\documentstyle[psfig,12pt]{article}
\def\virial{$\vert 2T/W\vert$}

\newcommand{\re}{\par\hangindent=0.5cm\hangafter=1\noindent}

\begin{document}
\begin{center}
{\Large \bf Growth of Velocity Dispersions for\\
\vspace{2mm}
Collapsing Spherical Stellar Systems}
\vspace*{5mm}

Shunsuke H{\sc ozumi}\\
{\it Faculty of Education, Shiga University, 2-5-1 Hiratsu, Otsu 520\\
E-mail: hozumi@sue.shiga-u.ac.jp}

Takao F{\sc ujiwara}\\
{\it Kyoto City University of Arts, Nishikyo-ku, Kyoto 610-11}\\

and

Yukitoshi K{\sc an-ya}\\
{\it Department of Physics, Faculty of Science, Kyoto University,\\
Sakyo-ku, Kyoto 606}\\

(Received 1996 February 5; accepted )

\vspace*{5mm}
{\bf Abstract}
\end{center}

First, we have ensured that spherical nonrotating collisionless systems
collapse with almost retaining spherical configurations during initial
contraction phases even if they are allowed to collapse three-dimensionally.
Next, on the assumption of spherical symmetry, we examine the evolution of
velocity dispersions with collapse for the systems which have uniform or
power-law density profiles with Maxwellian velocity distributions by
integrating the collisionless Boltzmann equation directly.  The results
show that as far as the initial contraction phases are concerned,
the radial velocity dispersion never grows faster than the tangential
velocity dispersion except at small radii where the nearly isothermal
nature remains, irrespective of the density profiles and virial ratios.
This implies that velocity anisotropy as an initial condition should be
a poor indicator for the radial orbit instability.  The growing behavior
of the velocity dispersions is briefly discussed from the viewpoint that
phase space density is conserved in collisionless systems.


\vspace*{5mm}
\begin{flushleft}
{\bf 1. Introduction}
\end{flushleft}

Since van Albada (1982) demonstrated that collapse of stellar systems with
small initial virial ratios leads to the end products whose density profiles
are well-described by the de Vaucouleurs $R^{1/4}$ law in projection, much
attention has been paid to cold dissipationless collapse from the standpoint
of the formation of elliptical galaxies.  In addition, it has been found
from collapse simulations (Polyachenko 1981, 1992; Merritt and Aguilar
1985; Barnes et al. 1986; Aguilar and Merrit 1990; Londrillo et al. 1991;
Udry 1993) that cold enough systems are deformed into triaxial configurations
by the radial orbit instability (ROI).  On the other hand, observations
suggest that elliptical galaxies are triaxial systems supported by
anisotropic velocity dispersions (Franx et al. 1991).  Putting these things
together, we may favor the view that the ROI accompanied by cold collapse
can naturally explain the observed properties of elliptical galaxies.

In spite of the importance of the ROI, its physical mechanism remains still
uncertain.  Many workers have searched for the criterion of the ROI from
the initial conditions adopted.  In particular, velocity anisotropy is
considered a key ingredient because the ROI could be a kind of Jeans
instability in the tangential direction as stated by \mbox{Merritt} (1987).
Although the ROI appears to occur in systems with larger velocity
dispersion in the radial direction than in the tangential direction,
Udry (1993) has reported that for centrally concentrated density profiles,
triaxial systems formed through the ROI even in the initial anisotropic
models characterized by 2$T_{\rm rad}/T_{\bot}\ll 1$, where $T_{\rm rad}$
and $T_{\bot}$ are the kinetic energies in radial and tangential motions,
respectively.  In retrospect, velocity anisotropy as an initial condition
may have nothing to do with the ROI.  In fact, numerical studies carried
out so far demonstrate that collapsing spherical nonrotating systems remain
spherical until around the maximum collapse times at which the virial
ratios become maximum (see e.g., Figure 3 of Aguilar and Merritt 1990).
Nevertheless, no one has argued the velocity anisotropy just before the
onset of the ROI.

These studies mentioned above indicate that we still lack the exact
knowledge of how the velocity dispersions of stellar systems, initially
far from equilibrium, evolve with collapse.  Recently, Kan-ya et al.
(1995) have shown that for spherically symmetric stellar systems the
tangential velocity dispersion grows faster than the radial one as far
as the initial contraction phases are concerned.  Unfortunately, however,
their analyses are based on such a cold approximation that the system has
nearly zero velocity dispersions initially.  In reality, stellar systems
would form with some degree of velocity dispersion.  If the system with
a power-law density profile has velocity dispersion at its birth, individual
mass shells can overlap with one another during the collapse, which is then
likely to affect the growing behavior of the velocity dispersions.  Aside
from the cold approximation, they analyzed the systems on the basis of
spherical symmetry.  Consequently, we cannot apply their results directly to
real galaxies which would have collapsed three-dimensionally.  Therefore,
we need to unravel the collapsing behavior of stellar systems at early
stages, and then investigate the growing behavior of the velocity dispersions
which they have from the beginning.

First, in section 2, we ensure that spherical stellar systems evolve
without deviating from spherical configurations at the early stages of
gravitational collapse.  Next, in section 3, on the assumption of spherical
symmetry, we study the growing behavior of velocity dispersions for the
systems which have uniform or power-law density profiles.  Discussion and
conclusions are presented in section 4.

\vspace*{5mm}
\begin{flushleft}
{\bf 2. Collapsing Behavior of Spherical Systems}
\end{flushleft}

We first study how spherical stellar systems, initially far from equilibrium,
collapse three-dimensionally at the early stages of contraction.  Above all,
the time evolution of axis ratios, $b/a$ and $c/a$, is presented, where $a,
b$, and $c$ are the major, intermediate, and minor axes, respectively.

We use the self-consistent field (SCF) method, termed by Hernquist and
Ostriker (1992), to study the collapsing behavior of spherical stellar
systems.  Since the SCF method requires no introduction of a softening
length, it is suitable to cold collapse simulations in which the
modification of a central force field through a softening length can have
a large effect on the global dynamics.  In addition, the reliability of
the SCF method is demonstrated by Hozumi and Hernquist (1995) who applied
it to spherically symmetric systems.  In the present simulations, we no
longer restrict ourselves to spherical symmetry.  Thus, the angular
dependence of the density and potential as well as the radial one is
expanded in the SCF code.  Since collapse simulations reveal that
density profiles of the end products resemble galaxies obeying the de
Vaucouleurs $R^{1/4}$ law in projection, we choose the basis set proposed
by Hernquist and Ostriker (1992) on the ground that its lowest order
members accurately describe spheroidal objects like elliptical galaxies
(\mbox{Hernquist} 1990).  We adopt $n_{\rm max}=16$ and $l_{\rm max}=%
m_{\rm max}=6$, where $n_{\max}$ is the maximum number of radial expansion
coefficients and $l_{\rm max}$ and $m_{\rm max}$ are the maximum numbers
of angular expansion coefficients.  We employ $N=100,000$ particles of
equal mass.  The equations of motion are integrated in Cartesian coordinates
using a time-centered leapfrog algorithm (e.g., Press et al. 1986).

The initial conditions consist of those spherical systems of total mass $M$
and radius $R_0$ which have density profiles of $\rho(r) \propto r^{-n}$,
where $n$ is chosen to be 0 and 1.  The velocity distributions are Maxwellian
with the dispersions determined from the virial ratio, \virial =$10^{-1.5}$,
where $T$ and $W$ denote the kinetic and gravitational energies, respectively,
although the velocities are assigned so that the initial data do not include
escapers. The unit of mass and the gravitational constant, $G$, are taken so
that $M=1$ and $G=1$.  The unit of length is determined from the relation
such that \virial $\times R_0=1$.  Then, we choose the length scale of the
basis functions to be 2.

\begin{center}
\leavevmode\psfig{file=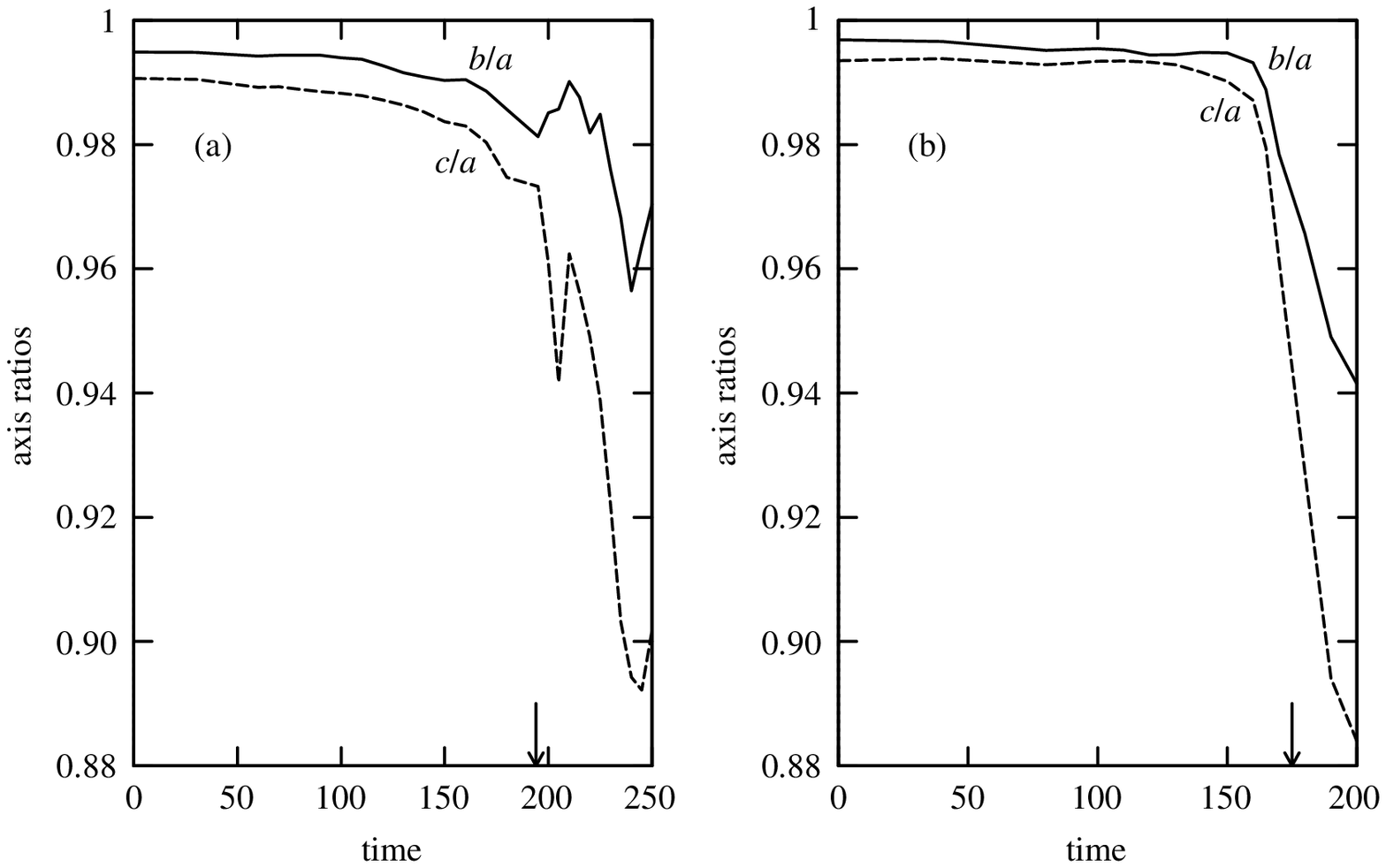,height=6cm,angle=-90}
\end{center}

\begin{quotation}
{\re \footnotesize
Fig.~1. Evolution of the axis ratios for (a) the uniform density sphere
and (b) the $\rho \propto r^{-1}$ model.  The virial ratios are chosen to
be $10^{-1.5}$.  The solid lines show $b/a$ and the dashed lines $c/a$,
where $a$, $b$, and $c$ denote the major, intermediate, and minor axes,
respectively.  The vertical arrows indicate the maximum collapse times.
}
\end{quotation}

In figures 1a and 1b, we present the time evolution of the axis ratios, $b/a$
and $c/a$ for $n=0$ and $1$, respectively.  The axis ratios are computed from
the principal moments of inertia for the most bound 80\% of the particles.  We
stop the simulations just after the maximum collapse times because we pay
attention only to the early stages of gravitational collapse.  According to
Cannizzo and Hollister (1992), the final elongation becomes larger as the
initial density profile is more centrally concentrated provided that the
initial virial ratio is fixed.  However, as far as the initial contraction
phases are concerned, we can see that the systems approach the maximum collapse
times with almost retaining spherical configurations, irrespective of the
initial density profiles.  In fact, figure 1b shows that the axis ratios
change little from the initial values of $b/a=0.997$ and $c/a=0.993$ during
the early contraction times (from $t=0$ to $t \approx 170$) even for the $n=1$
density profile model.  The small deviations of $b/a$ and $c/a$ from unity
before the maximum collapse phases are, in large measure, considered
root-mean-square fluctuations coming from finite numbers of particles used
in the simulations.  We conclude, therefore, that spherical stellar systems
which have uniform or power-law density profiles do not deviate considerably
from spherical configurations during the early stages of contraction even
if they are allowed to collapse three-dimensionally.

\vspace*{5mm}
\begin{flushleft}
{\bf 3. Growing Behavior of Velocity Dispersions}
\end{flushleft}

We have demonstrated in section 2 that spherical stellar systems collapse
symmetrically in configuration to a considerable degree at the early stages
of contraction.  Thus, spherical symmetry can be a good assumption when we
study the growing behavior of velocity dispersions for such systems as long
as their evolution stays in the initial contraction phases.

\vspace*{5mm}
\begin{flushleft}
{\it 3}.{\it 1}. {\it Analytical Predictions}
\end{flushleft}

Before proceeding into numerical simulations, we summarize the growing
behavior of the velocity dispersions predicted from analytical results.
In figure 2, we reproduce the evolution of the velocity dispersions for
spherical symmetric systems on the basis of Kan-ya et al.'s (1995) analyses.
It can be noticed from figure 2a that for the uniform density sphere
the radial velocity dispersion grows exactly in the same manner as the
tangential one and that the growth of both the dispersions is independent
of radius.  Figures 2b and 2c show that for the power-law density spheres
the radial velocity dispersions decrease after some collapse time, while
the tangential velocity dispersions continue to increase with time at all
radii.  In particular, the radial velocity dispersions decrease inward
radially after some collapse time for the $\rho \propto r^{-0.5}$ model,
and always decrease inward radially for the $\rho \propto r^{-1}$ model.
Of course, this behavior of the radial velocity dispersion comes from
Kan-ya et al.'s (1995) assumption that the system can continue to
contract infinitely.  Furthermore, they neglect the existence of the outer
edges.  Therefore, their analyses would apply to some intermediate radii
of the systems which have finite extent and some degree of velocity
dispersion from the beginning.

\begin{center}
\leavevmode\psfig{file=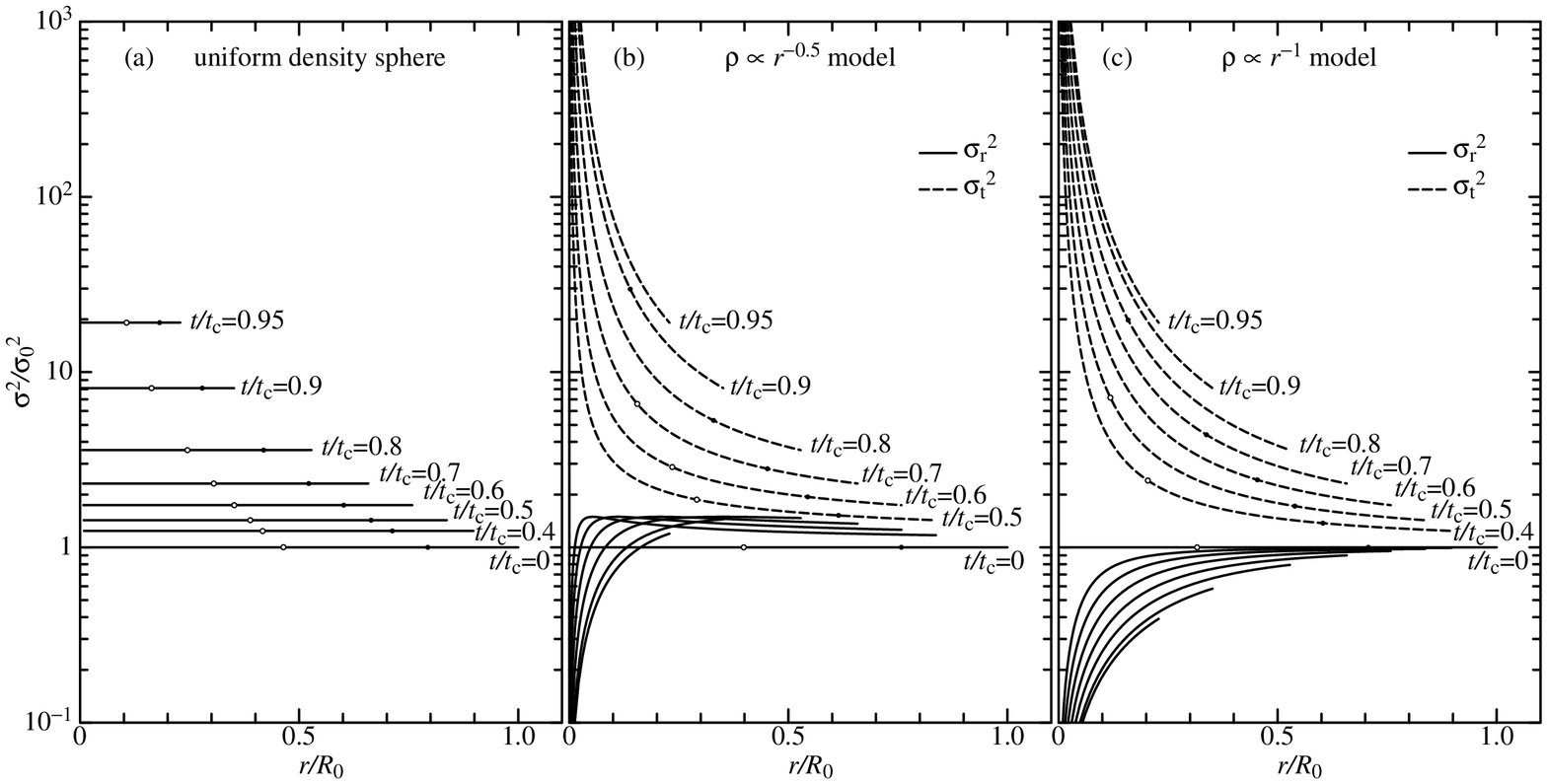,height=6.5cm,angle=-90}
\end{center}

\begin{quotation}
{\re \footnotesize
Fig.~2. Evolution of the velocity dispersions on the assumption that the
effects of velocity dispersions are small enough, for (a) the uniform density
sphere, (b) the $\rho \propto r^{-0.5}$ model, and (c) the $\rho
\propto r^{-1}$ model.  The solid lines show the square of the radial
velocity dispersion $\sigma_{\rm r}^2$, and the dashed lines that of
the tangential velocity dispersion $\sigma_{\rm t}^2$.  Note that the
evolution of the radial velocity dispersion is exactly the same as that
of the tangential velocity dispersion for the uniform density sphere.
The velocity dispersions and the radius are normalized with the initial
value $\sigma_0^2$ and the cutoff radius $R_0$, respectively.  The circles
are added on the $\sigma_{\rm t}^2$ lines to indicate the evolutionary
stages of the collapse: the filled circle is the half mass radius and the
open circle the radius within which the system contains 10 \% of the total
mass.  The time $t_{\rm c}$ is the collapse time at the cutoff radius $R_0$.
}
\end{quotation}

In summary, we can see that the growing rate of the radial velocity
dispersion never exceeds that of the tangential one with a cold
approximation.  In particular, for the power-law density spheres, the
ratio of the radial to tangential velocity dispersions becomes smaller
as time proceeds.  In addition, the decrease of the ratio is accelerated
as the density gradient becomes steeper.

\vspace*{5mm}
\begin{flushleft}
{\it 3}.{\it 2}. {\it Numerical Results}
\end{flushleft}

Now that spherical symmetry is adopted, we use a phase-space method
developed by Fujiwara (1983) because it gives desirable results
insusceptible to random fluctuations.  This method enables us to obtain a
smooth distribution of velocity dispersions along radius.  We employ
$(N_r, N_u, N_j)=(150, 151, 80)$, where $N_r, N_u$, and $N_j$ are the
numbers of mesh points along radius, radial velocity, and angular momentum,
respectively.  The units of mass, $M$, and the gravitational constant, $G$,
are taken so that $M=1$ and $G=1$.  The unit of length is determined from
the relation such that \virial $\times R_0=1$ as in section 2.

We examine the time evolution of the velocity dispersions by gravitational
collapse for density profiles of $\rho(r) \propto r^{-n}$ with Maxwellian
velocity distributions, where $n$=0, 1/2, and 1 are chosen, on the
assumption of spherical symmetry.  The initial virial ratios are taken
to be $10^{-1.5}$, $10^{-1}$, and $10^{-0.5}$ for each value of $n$.

In figure 3, the time evolution of $\sigma_{\rm r}^2$ and $\sigma_{\rm t}^2$
is shown for the uniform density spheres, i.e., $n$=0, where $\sigma_{\rm r}$
and $\sigma_{\rm t}$ are the radial and tangential velocity dispersions,
respectively.  It can be noticed that the evolution of $\sigma_{\rm r}^2$
and $\sigma_{\rm t}^2$ is very similar to one another, regardless of the
initial virial ratios.  That is, both the velocity dispersions grow
at the same rate within some radius, which means that the systems keep
the isothermal nature there as indicated by the analytical predictions.
The outer regions, however, show the deviation from them.  This is because
the velocity dispersions which the system has from the beginning operate to
blur the outer edge: the spread of the radial velocity becomes smaller
outward to the edge.  In this way, the isothermal nature disappears with
time from outside.  This effect is more prominent for the larger initial
virial ratios. Therefore, the coldest model shows the abrupt drop in the
radial velocity dispersion (figure 3a), while the other models show the
gentle decrease of it (figures 3b and 3c).

\begin{center}
\leavevmode\psfig{file=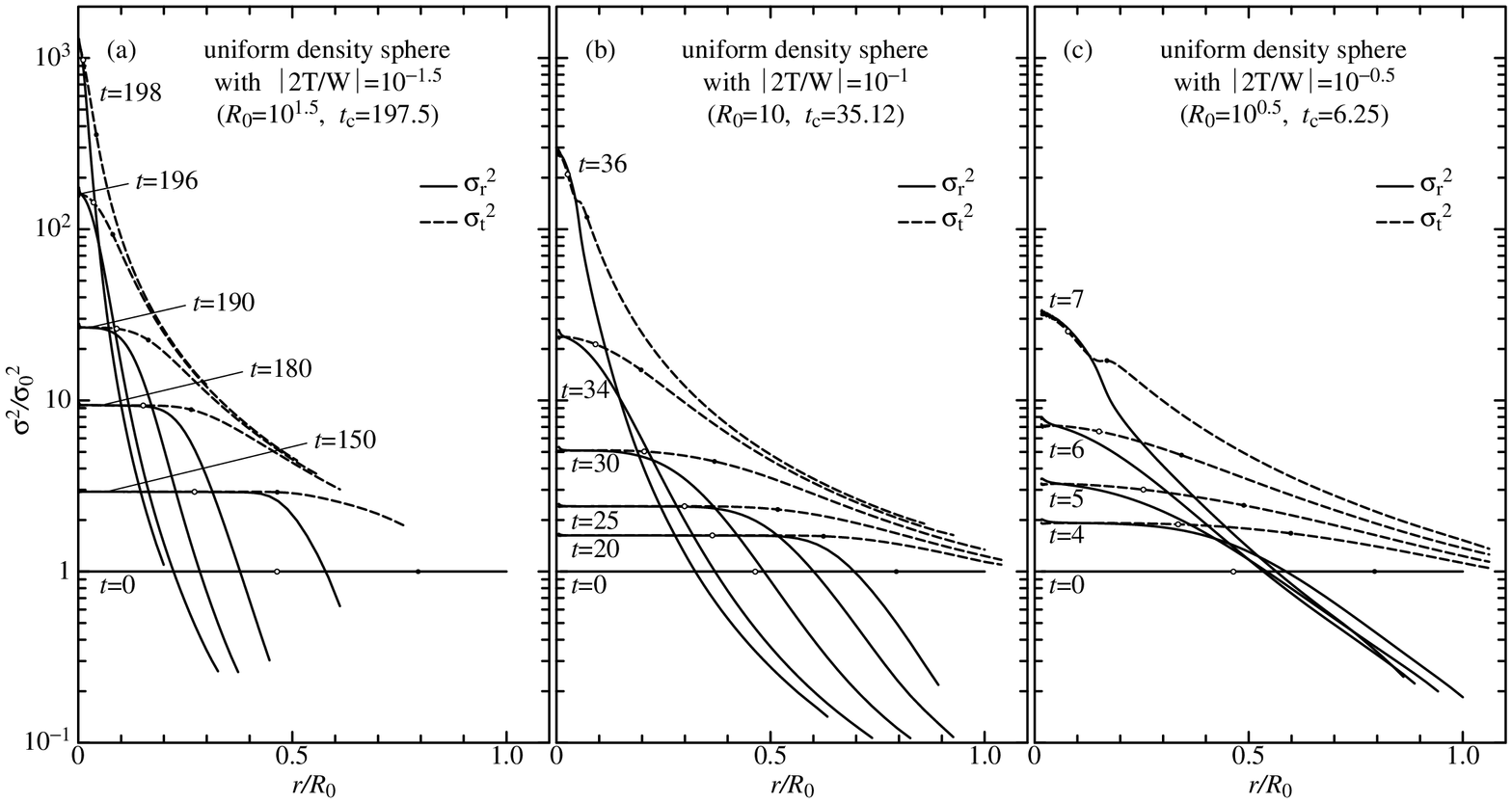,height=6.5cm,angle=-90}
\end{center}

\begin{quotation}
{\re \footnotesize
Fig.~3. Evolution of the velocity dispersions for the uniform density spheres
with initial virial ratios (a) \virial $=10^{-1.5}$, (b) \virial $=10^{-1}$,
and (c) \virial $=10^{-0.5}$.  The solid lines show the square of the
radial velocity dispersion $\sigma_{\rm r}^2$ and the dashed lines that of
the tangential velocity dispersion $\sigma_{\rm t}^2$.  The velocity
dispersions and the radius are normalized with the initial value
$\sigma_0^2$ and the cutoff radius $R_0$, respectively.  The filled circle
is the half mass radius and the open circle the radius within which the
system contains 10 \% of the total mass.  The time $t_{\rm c}$ is the
collapse time at the cutoff radius $R_0$ with no initial velocity dispersion.
}
\end{quotation}

In figures 4 and 5, we present the time evolution of $\sigma_{\rm r}^2$
and $\sigma_{\rm t}^2$ for the $\rho \propto r^{-0.5}$ and $\rho \propto
r^{-1}$ models, respectively.  Now, let us focus on figures 4a and 5a.  If
we pay attention to the intermediate radii, the tendency of the evolution
of both the velocity dispersions agrees roughly with that predicted by the
analytical results (see figures 2b and 2c).  In the central region,
however, the nearly isothermal evolution can be seen, which differs from
the analytical results.  This is because the velocity dispersions prevent
the system from contracting infinitely.  In fact, the density near the
central region was kept approximately uniform throughout the contracting
phase.  Thus, the evolution of the central part resembles that of the
uniform density spheres: both the radial and tangential velocity dispersions
increase in the same manner as the system collapses.  The deviation from
the analytical predictions also appears near the outer edges.  The same
reason as described for the uniform

\begin{center}
\leavevmode\psfig{file=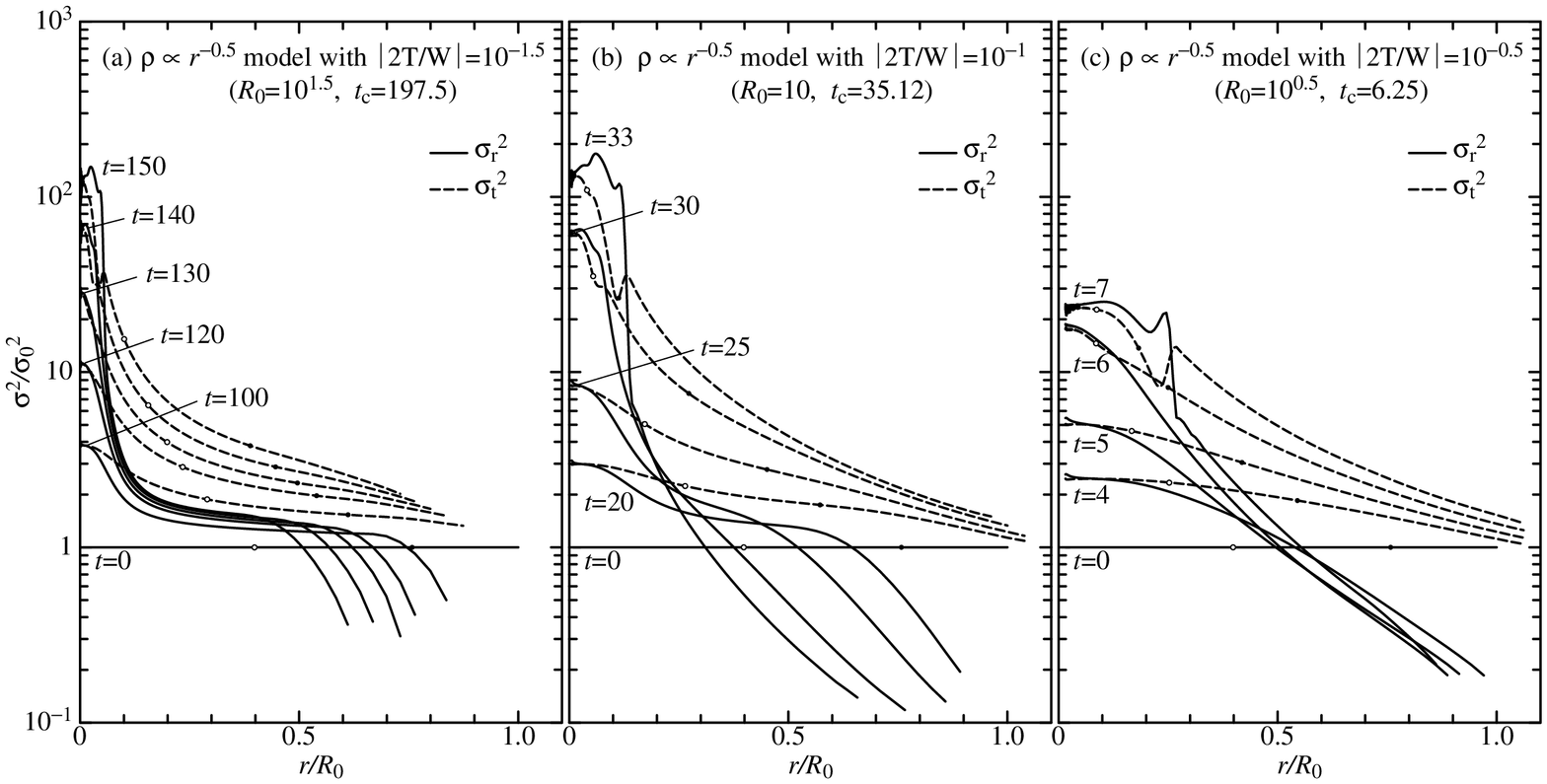,height=6.5cm,angle=-90}
\end{center}

\begin{quotation}
{\re \footnotesize
Fig.~4. Evolution of the velocity dispersions for the $\rho \propto r^{-0.5}$
models with initial virial ratios (a) \virial $=10^{-1.5}$, (b) \virial
$=10^{-1}$, and (c) \virial $=10^{-0.5}$.  The notations and their meanings
are the same as in figure 2.
}
\end{quotation}

\vspace*{5mm}
\begin{center}
\leavevmode\psfig{file=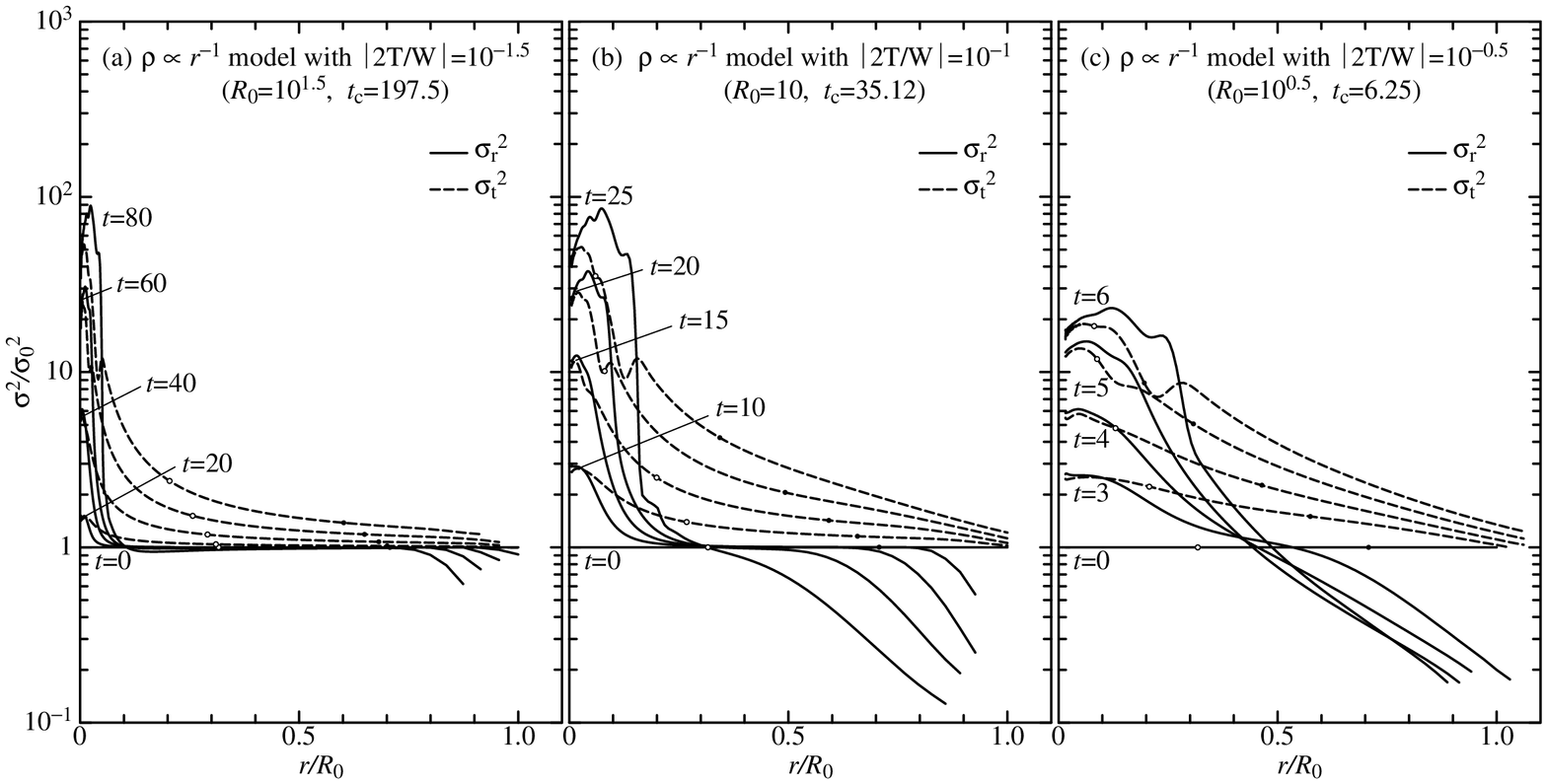,height=6.5cm,angle=-90}
\end{center}

\begin{quotation}
{\re \footnotesize
Fig.~5. Evolution of the velocity dispersions for the $\rho \propto r^{-1}$
models with initial virial ratios (a) \virial $=10^{-1.5}$, (b) \virial
$=10^{-1}$, and (c) \virial $=10^{-0.5}$.  The notations and their meanings
are the same as in figure 2.
}
\end{quotation}
density spheres can be applied to the
decrease of the radial velocity dispersion near the edges.  The larger
amount of velocity dispersion does not change largely the evolution of
$\sigma_{\rm r}^2$ and $\sigma_{\rm t}^2$ as can be seen from figures 4b,
4c, 5b, and 5c.

\vspace*{5mm}
\begin{flushleft}
{\bf 4. Discussion and Conclusions}
\end{flushleft}

We have found that spherical nonrotating systems keep their spherical
configurations during the collapse at the early stages of contraction.
We thus carried out collapse simulations on the assumption of spherical
symmetry to examine how the velocity dispersions of spherical stellar
systems evolve with collapse.  Then, we have demonstrated that the radial
velocity dispersion never grows faster than the tangential velocity
dispersion except at small radii where the nearly isothermal nature is
retained up to around the maximum collapse times, irrespective of the
initial density profiles and virial ratios.

Our results are basically consistent with those of Kan-ya et al. (1995),
although they assume that initial systems have nearly zero velocity
dispersions.  Thus, we can see that in general the system just before the
onset of the ROI has larger velocity dispersion in the tangential direction
than in the radial direction, which might be different from the situation
that we envisage for the occurrence of the ROI.  However, these findings can
be understood qualitatively with relative ease as follows.  If the initial
density distribution is uniform, the collapse time of the system is
independent of radius, so that the conservation of phase space density
results in $\sigma_{\rm R} \times R=const$., where $\sigma_{\rm R}$ and
$R$ denote the characteristic radial velocity dispersion and radius of
the system, respectively.  In addition, the conservation of angular momentum
of each star leads to $\sigma_{\rm T} \times R=const$., where $\sigma_{\rm T}$
is the characteristic one-dimensional tangential velocity dispersion.
Then, the ratio, $\sigma_{\rm R}/\sigma_{\rm T}$, remains unity during
the early stages of contraction when the system starts with an isotropic
velocity distribution.  On the other hand, if the initial density
distribution is power-law such as $\rho \propto r^{-n} (n > 0)$, the stars
of inner parts fall first and then those of outer parts follow.  Consequently,
the width of the radial velocity spread for the power-law density profiles
is narrower than that for the uniform density profiles while $\sigma_{\rm T}
\times R=const$.\ still holds.  It may be helpful to remind that the collapse
time of the outer edge for the power-law density spheres is the same as that
for the uniform density spheres.  As a result, the ratio,
$\sigma_{\rm R}/\sigma_{\rm T}$, continues to decrease from its initial
value of unity with time if the system has a power-law density profile and
an isotropic velocity distribution at the beginning.

In the numerical simulations, the deviation of the evolution of the velocity
dispersions from the analytical results emerges at small radii for the
power-law density profiles and near the edge of the system for the uniform
and power-law density profiles.  Since the system has finite extent, the
velocity difference of the phase space elements makes the radial velocity
dispersion smaller near the edge as collapse proceeds.  On the other hand,
the velocity dispersions can save the system from contracting infinitely.
Then, the conservation of phase space density again means the increase of
the radial velocity dispersion as the system contracts.  Concerning the
isothermal nature at small radii, we point out that the density near the
central region was kept approximately uniform with collapse.

We can now explain why the collapse simulations thus far have shown that
spherical systems approached the maximum collapse phases with retaining
their initial shapes.  This is because the tangential velocity dispersion
remains equal to or overwhelms the radial one as collapse proceeds.
Conversely, the predominance of the growth of the tangential velocity
dispersion prevents the system from being deformed into elongated
configurations at early collapse phases.  It is thus suggested that
the onset of the ROI is determined from the physical conditions at least
just before or around the maximum collapse phase.  Even though we start
with the condition that $2T_{\rm rad}/T_{\bot} > 1$, the growth of the
tangential velocity dispersion can easily catch up with or exceed that
of the radial velocity dispersion if the system has a power-law density
distribution.  Since we have found that the criterion of the ROI will not
depend strongly on the initial velocity anisotropy, it is no wonder that
small values such that $2T_{\rm rad}/T_{\bot}\ll 1$ lead to triaxial end
products as reported by Udry (1993).  Therefore, we conclude that initial
anisotropy parameters like $2T_{\rm rad}/T_{\bot}$ should be a poor
indicator for the ROI.

\vspace*{5mm}
We are grateful to Professors S. Kato and S. Inagaki for their careful
reading of the manuscript and helpful comments on its contents.

\vspace*{5mm}
\begin{flushleft}
{\bf References}
\end{flushleft}

\def\refpar{\parindent=0pt%
            \hangafter=1 \hangindent=3ex}

\noindent
Aguilar L.A., Merritt D.R. 1990, ApJ 265, 33

\noindent
Barnes J.E., Goodman J., Hut P. 1986, ApJ 300, 122

\noindent
Cannizzo J.K., Hollister T.C. 1992, ApJ 400, 58

\noindent
Franx M., Illingworth G.D., de Zeeuw P.T. 1991, ApJ 383, 112

\noindent
Fujiwara T. 1983, PASJ 35, 547

\noindent
Hernquist L. 1990, ApJ 356, 359

\noindent
Hernquist L., Ostriker J.P. 1992, ApJ 386, 375

\noindent
Hozumi S., Hernquist L. 1995, ApJ 440, 60

\refpar
Kan-ya Y., Sasaki M., Tsuchiya T., Gouda N. 1995, Submitted to Prog.
Theor.Phys.

\noindent
Londrillo P., Messina A., Stiavelli M. 1991, MNRAS 250, 54

\refpar
Merritt D.R. 1987, in Structure and Dynamics of Elliptical Galaxies,
IAU Symp. No. 127, ed. P.T. de Zeeuw (Reidel, Dordrecht), p315

\noindent
Merritt D.R., Aguilar L.A. 1985, MNRAS 217, 787

\noindent
Polyachenko V.L. 1981, SvA L7, 79

\noindent
Polyachenko V.L. 1992, SvA 36, 482

\refpar
Press W.H., Flannery B.P., Teukolsky S.A., Vetterling W.T. 1986,
in Numerical Recipes: The Art of Scientific Computing
(Cambridge University Press, Cambridge)

\noindent
Udry, S. 1993, A\&A 268, 35

\noindent
van Albada T.S. 1982, MNRAS 201, 939

\end{document}